\documentclass[aps,showpacs,preprintnumbers,amsmath, amssymb]{revtex4}

\usepackage{float}
\usepackage{graphics,epsfig}
\usepackage{graphicx}
\usepackage{dcolumn}
\usepackage{bm}

\begin{document}
\baselineskip=0.8 cm

\title{{\bf Scalar field configurations supported by charged compact reflecting stars in a curved spacetime}}
\author{Yan Peng$^{1}$\footnote{yanpengphy@163.com}}
\affiliation{\\$^{1}$ School of Mathematical Sciences, Qufu Normal University, Qufu, Shandong 273165, China}

\vspace*{0.2cm}
\begin{abstract}
\baselineskip=0.6 cm
\begin{center}
{\bf Abstract}
\end{center}

We study the system of static scalar fields coupled to charged compact reflecting stars
through both analytical and numerical methods.
We enclose the star in a box and our solutions are related to cases without box boundaries
when putting the box far away from the star.
We provide bottom and upper bounds for the radius of the scalar hairy compact reflecting star.
We obtain numerical scalar hairy star solutions satisfying boundary conditions
and find that the radius of the hairy star in a box is continuous in a range,
which is very different from cases without box boundaries where the radius is discrete in the range.
We also examine effects of the star charge and mass on the largest radius.

\end{abstract}

\maketitle
\newpage
\vspace*{0.2cm}

\section{Introduction}

The no-scalar-hair theorem is a famous physical
characteristic of black holes \cite{Bekenstein,Chase,Ruffini-1}.
It was found that the static massive scalar fields cannot
exist in asymptotically flat black holes, for references see \cite{Hod-1}-\cite{Brihaye}
and a review see \cite{Bekenstein-1}.
This property is usually attributed to the
fact that the horizon of a classical black hole irreversibly absorbs matter and radiation fields.
Along this line, one naturally want to know whether this no scalar hair behavior is a
unique property of black holes. So it is interesting to explore possible similar no scalar hair theorem
in other horizonless curved spacetimes.

Lately, hod found a no-scalar-hair theorem for asymptotically flat horizonless neutral compact reflecting
stars with a single massive scalar field and specific types of the potential \cite{Hod-6}.
Bhattacharjee and Sudipta further extended the discussion to spacetimes with a positive cosmological constant \cite{Bhattacharjee}.
In fact, the no scalar hair behavior also exists for massless
scalar field nonminimally coupled to gravity on the  neutral compact reflecting star background \cite{Hod-7}.
Recently, scalar field configurations were constructed in the charged compact reflecting shell
where the star charge and mass can be neglected compared to the star radius \cite{Hod-8,Hod-9}.
With analytical methods, the physical properties of the asymptotically
flat composed star-field configurations were also analyzed in \cite{Hod-10}.
In particular, this work derived a remarkably compact analytical formula for
the discrete spectrum of star radii.
Along this line, it is interesting to extend the discussion by relaxing the condition that
star radii are much larger than the star charge and mass.

On the other side, a simple way to invade the black hole
no-scalar-hair theorem is adding a reflecting box boundary.
It should be emphasized that the boundary conditions imposed by a box are different
from the familiar boundary conditions of asymptotically flat spacetimes.
In fact, it was found that the low frequency scalar field perturbation can trigger
superradiant instability of the charged black hole in a box
and the nonlinear dynamical evolution can form a quasi-local
hairy black hole \cite{Carlos, Dolan, Supakchai, Nicolas}.
From thermodynamical aspects, Pallab and other authors showed that
there are stable asymptotically flat hairy black holes in a box
invading no-hair-theorem of black holes \cite{Pallab Basu}.
It was believed that the box boundary could play a role of the infinity
potential to make the fields bounce back and condense around the black hole.
Along this line, it is interesting to extend the discussion of
scalar field configurations supported by a compact
reflecting star through including an additional box boundary
and also compare mathematical structures between gravities
without box boundaries and models in a box .

The next sections are planed as follows. In section II,
we introduce the model of a charged compact reflecting star coupled to a scalar field.
In part A of section III, we provide bounds for the radius of the scalar hairy star.
In part B of section III, we obtain radius of the hairy star and explore effects of
parameters on the largest radius. And in part C of section III, we also
carry out an analytical study of the system in the limit that star charge and mass can be neglected.
We will summarize our main results in the last section.

\section{Equations of motion and boundary conditions}

We consider the system of a scalar field and a compact reflecting
star enclosed in a time-like reflecting box at $r=r_{b}$
in the four dimensional asymptotically flat gravity.
When $r_{b}\rightarrow\infty$, we go back to the case without box boundaries.
We also define the radial coordinate $r=r_{s}$ as the radius of the compact star.
And the corresponding Lagrange density is given by
\begin{eqnarray}\label{lagrange-1}
\mathcal{L}=-\frac{1}{4}F^{MN}F_{MN}-|\nabla_{\mu} \psi-q A_{\mu}\psi|^{2}-\mu^{2}\psi^{2},
\end{eqnarray}
where q and $\mu$ are the charge and mass of the scalar field $\psi(r)$ respectively.
And $A_{\mu}$ stands for the ordinary Maxwell field.

Using the Schwarzschild coordinates, the line element of the
spherically symmetric star can be expressed in the form \cite{Chandrasekhar}
\begin{eqnarray}\label{AdSBH}
ds^{2}&=&-(1-\frac{2M}{r}+\frac{Q^2}{r^2})dt^{2}+\frac{dr^{2}}{1-\frac{2M}{r}+\frac{Q^2}{r^2}}+r^{2}(d\theta^{2}+sin^{2}\theta d\varphi^{2}).
\end{eqnarray}
where $M$ is the mass of the star and $Q$ is the charge of the star.
In this paper, we only study the case of $M\geqslant Q$. Since the spacetime is regular, we also assume that
$r_{s}>M+\sqrt{M^2-Q^2}$. And the Maxwell field with only the nonzero $tt$ component is $A_{\mu}=-\frac{Q}{r}dt$.

For simplicity, we study the scalar field with only radial dependence in the form $\psi=\psi(r)$.
From above assumptions, we obtain equations of motion as
\begin{eqnarray}\label{BHg}
\psi''+(\frac{2}{r}+\frac{g'}{g})\psi'+(\frac{q^2Q^2}{r^2g^2}-\frac{\mu^2}{g})\psi=0,
\end{eqnarray}
with $g=1-\frac{2M}{r}+\frac{Q^2}{r^2}$.

In addition, we impose reflecting boundary conditions for the scalar field
at the surface of the compact star.
We also suppose that the time-like box boundary $r=r_{b}$ can reflect the scalar field back.
So the scalar field vanishes at the boundaries as
\begin{eqnarray}\label{InfBH}
&&\psi(r_{s})=0,~~~~~~~~~\psi(r_{b})=0.
\end{eqnarray}

\section{Scalar field configurations in charged compact reflecting stars}

\subsection{Bounds for the radius of the scalar hairy compact star}

Defining the new radial function $\tilde{\psi}=\sqrt{r}\psi$,
one obtains the differential equation
\begin{eqnarray}\label{BHg}
r^2\tilde{\psi}''+(r+\frac{r^2g'}{g})\tilde{\psi}'+(-\frac{1}{4}-\frac{rg'}{2g}+\frac{q^2Q^2}{g^2}-\frac{\mu^2r^2}{g})\tilde{\psi}=0,
\end{eqnarray}
with $g=1-\frac{2M}{r}+\frac{Q^2}{r^2}$.

According to the boundary conditions (4), one deduce that
\begin{eqnarray}\label{InfBH}
&&\tilde{\psi}(r_{s})=0,~~~~~~~~~\tilde{\psi}(r_{b})=0.
\end{eqnarray}

The function $\tilde{\psi}$ must have (at least) one extremum point $r=r_{peak}$
between the surface $r_{s}$ of the reflecting star and the box boundary $r_{b}$ (including cases of $r_{b}=\infty$).
At this extremum point, the scalar field is characterized by
\begin{eqnarray}\label{InfBH}
\{ \tilde{\psi}'=0~~~~and~~~~\tilde{\psi} \tilde{\psi}''\leqslant0\}~~~~for~~~~r=r_{peak}.
\end{eqnarray}

According to the relations (5) and (7), we arrive at the inequality
\begin{eqnarray}\label{BHg}
-\frac{1}{4}-\frac{rg'}{2g}+\frac{q^2Q^2}{g^2}-\frac{\mu^2r^2}{g}\geqslant0~~~for~~~r=r_{peak}.
\end{eqnarray}

Then we have
\begin{eqnarray}\label{BHg}
\mu^2r^2g\leqslant q^2Q^2-\frac{rgg'}{2}-\frac{1}{4}g^2~~~for~~~r=r_{peak}.
\end{eqnarray}

Since $r\geqslant r_{s}> M+\sqrt{M^2-Q^2}\geqslant M \geqslant Q$, we have
\begin{eqnarray}\label{BHg}
g=1-\frac{2M}{r}+\frac{Q^2}{r^2}=\frac{1}{r^2}(r^2-2Mr+Q^2)=\frac{1}{r^2}[(r-M)^2-(M^2-Q^2)]\geqslant0,
\end{eqnarray}
\begin{eqnarray}\label{BHg}
rg'=r(1-\frac{2M}{r}+\frac{Q^2}{r^2})'=r(\frac{2M}{r^2}-\frac{2Q^2}{r^3})=\frac{2M}{r}(1-\frac{Q}{r}\frac{Q}{M})\geqslant 0
\end{eqnarray}
and
\begin{eqnarray}\label{BHg}
(r^2g)'=(r^2-2Mr+Q^2)'=2(r-M)\geqslant 0.
\end{eqnarray}

Then we arrive at
\begin{eqnarray}\label{BHg}
\mu^2r_{s}^2g(r_{s})\leqslant \mu^2r^2g(r)\leqslant q^2Q^2-\frac{rgg'}{2}-\frac{1}{4}g^2\leqslant q^2Q^2~~~for~~~r=r_{peak}.
\end{eqnarray}

According to (13), there is
\begin{eqnarray}\label{BHg}
\mu^2r_{s}^2g(r_{s})\leqslant  q^2Q^2.
\end{eqnarray}

Taking cognizance of the metric solutions, (14) can also be expressed as
\begin{eqnarray}\label{BHg}
\mu^2r_{s}^2(1-\frac{2M}{r_{s}}+\frac{Q^2}{r_{s}^2})\leqslant  q^2Q^2.
\end{eqnarray}

Then, we can transfer (15) into the form
\begin{eqnarray}\label{BHg}
(\mu r_{s})^2-(2\mu M)(\mu r_{s})+Q^2(\mu^2-q^2)\leqslant 0.
\end{eqnarray}

Then, we obtain bounds for the radius of the scalar hairy compact reflecting star as
\begin{eqnarray}\label{BHg}
\mu M+\sqrt{\mu^{2}(M^2-Q^2)} < \mu r_{s}\leqslant \mu M+\sqrt{\mu^{2}(M^2-Q^2)+q^2Q^2},
\end{eqnarray}
The bottom bound comes from the assumption that
the spacetime is regular or $r_{s}>M+\sqrt{M^2-Q^2}$
and the upper bound can be obtained from (16).
For a neutral scalar field with $q=0$, (17) shows that
the upper bound is behind a horizon meaning a no-hair-theorem for the neutral
scalar field in a charged reflecting star.
So it is the coupling $qQ$ makes the upper bound larger than the horizon critical points
and then the scalar hair can possibly exist in this regular spacetime.

\subsection{Scalar field configurations in a curved spacetime}

The scalar field configurations with charged reflecting stars were studied
in the limit of $Q,M\ll r_{s}$ \cite{Hod-8,Hod-9,Hod-10}.
In this part, we will extend the discussion by relaxing the condition $Q,M\ll r_{s}$.
We can simply set $\mu=1$ in the following calculation using the symmetry of the equation (3) in the form
\begin{eqnarray}\label{BHg}
r\rightarrow k r,~~~~ \mu\rightarrow \mu/k,~~~~ M\rightarrow k M,~~~~ Q\rightarrow k Q,~~~~ q\rightarrow q/k.
\end{eqnarray}
Around the star surface, the scalar field can be expanded as $\psi=\psi_{0}(r-r_{s})+\cdots$.
Since the scalar field equation is linear and homogeneous with respect to $\psi$,
we can fix $\psi_{0}=1$ and use the numerical shooting method
to integrate the equation from $r_{s}$ to the infinity to search for the proper $r_{b}$
satisfying the box boundary condition $\psi(r_{b})=0$.

Now, we show the numerical hairy compact star solutions
with $q=2$, $Q=4$ and $M=5$ in Fig. 1.
In the left panel, when we choose a radius $r_{s}=11.2$, the scalar field decreases
to be zero at $r_{b}\thickapprox13.76$ as the box boundary. In contrast, with a little larger radius $r_{s}=11.3$
in the middle panel, the box boundary can be fixed at a larger radius $r_{b}\thickapprox14.36$.
And for a radius $r_{s}=11.4$ in the right panel, the solutions behave
similar to $\psi\varpropto \frac{1}{r}e^{\mu r}$ far from the star and there is no points to impose the box boundary.
With more detailed calculations, we arrive at a critical value $R_{s}\thickapprox11.38808027313$, above which the reflecting star
cannot support scalar fields. And we can numerically
find a scalar hairy quasi-local reflecting star for radius $r_{s}\in(8,R_{s}]$
except discrete radius corresponding to $r_{b}=\infty$ as can be obtained in the right panel of Fig. 2.

\begin{figure}[h]
\includegraphics[width=155pt]{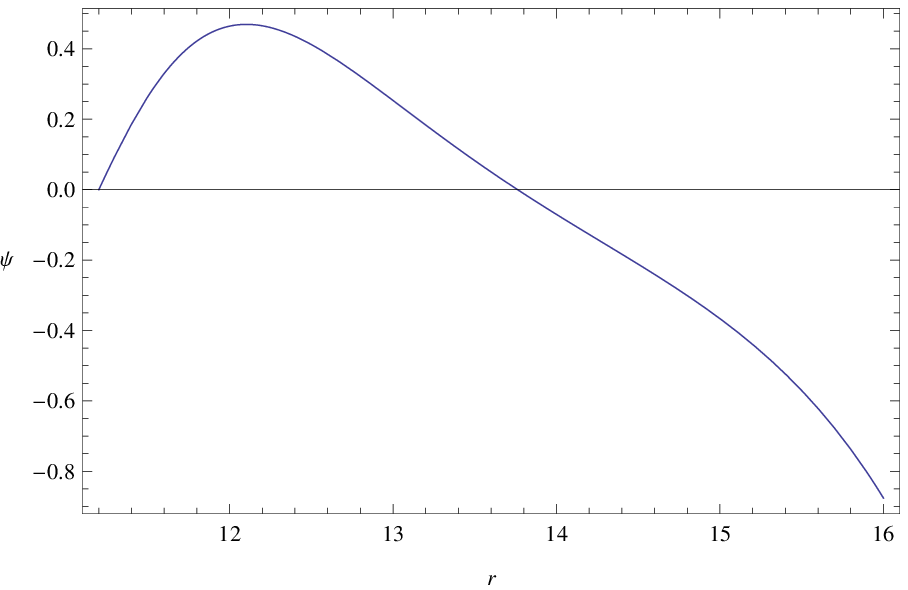}\
\includegraphics[width=155pt]{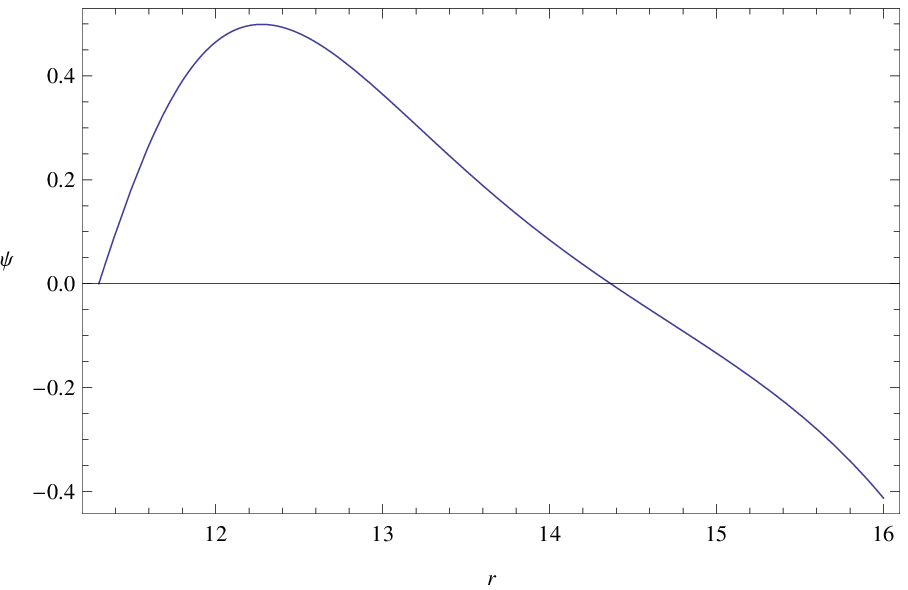}\
\includegraphics[width=155pt]{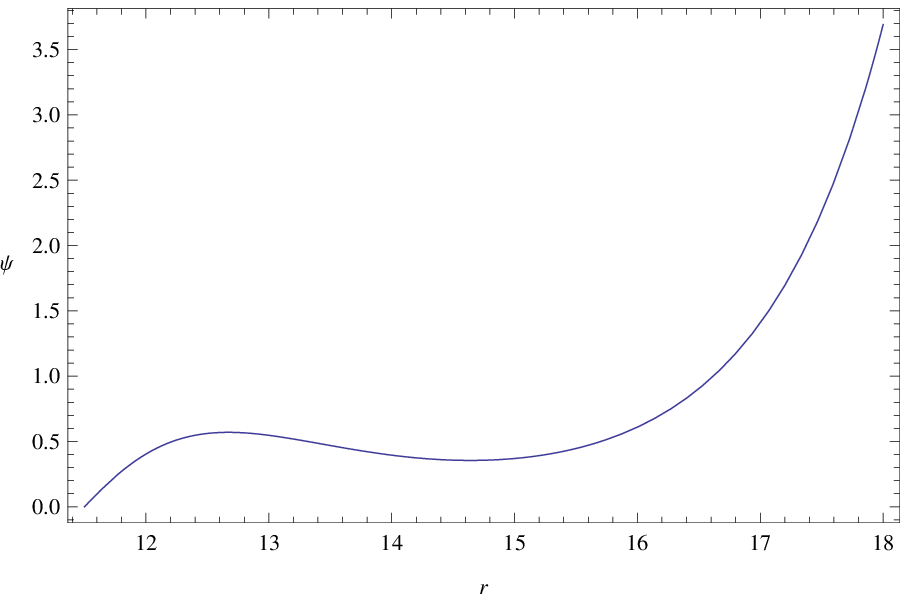}\
\caption{\label{EEntropySoliton} (Color online) We plot the function of $\psi$
with respect to the coordinate $r$ with $q=2$, $Q=4$ and $M=5$. The left panel shows the case
of $r_{s}=11.2$, the middle panel corresponds to the case of $r_{s}=11.3$
and the right panel represents the case of $r_{s}=11.4$.}
\end{figure}

We show behaviors of the scalar field with $q=2$, $Q=4$,
$M=5$ and the largest radius at $r_{s}=11.38808027313$ in Fig. 2.
We found that when $r_{s}\rightarrow R_{s}$,
the scalar field converges to a nonzero limit and
at the same time the boundary is $r_{b}\rightarrow \infty$.
So our numerical solutions return to cases without box boundaries as in the left panel of Fig. 2.
On the other side, the general solutions of equation (3) behave as
$\psi\thickapprox A\cdot\frac{1}{r}e^{-\mu r}+B\cdot\frac{1}{r}e^{\mu r}$ with $r\rightarrow \infty$.
We have carefully checked that B is negative in cases of $r_{s}<R_{s}$ and
there is $B>0$ for $r_{s}>R_{s}$, see also results in Fig. 1.
So a critical radius $r_{s}=R_{s}$ with $B=0$ should exist and that corresponds to a
scalar field with asymptotical behaviors $\psi\varpropto \frac{1}{r}e^{-\mu r}$ at the infinity
on a reflecting star background without box boundaries.

Integrating the equation from $r_{s}=11.38808027313$ to smaller
coordinates in the right panel of Fig. 2, we can obtain other possible
discrete radius of hairy stars without box boundaries.
It can be seen from the picture that the radius of the hairy star with $r_{b}=\infty$ can be
fixed at discrete points around $r_{s}\thickapprox11.39,~~10.26,~~9.58, \cdots$
between bounds about $8< r_{s}\leqslant 13.54$ according to (17).
We conclude that radius of scalar hairy reflecting stars without box boundaries in a curved spacetime
is discrete similar to cases in \cite{Hod-8,Hod-9,Hod-10}.

\begin{figure}[h]
\includegraphics[width=165pt]{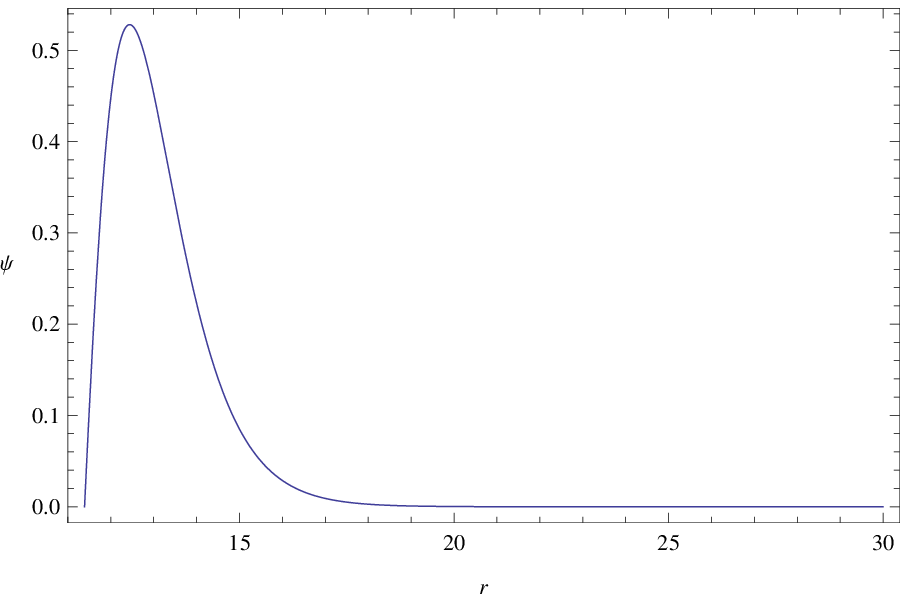}\
\includegraphics[width=168pt]{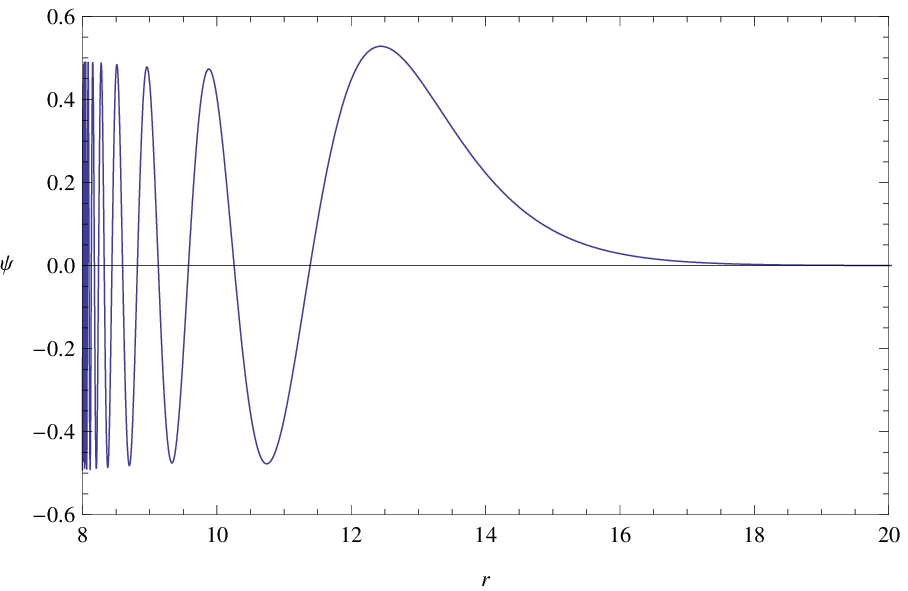}\
\caption{\label{EEntropySoliton} (Color online) We plot the function of $\psi$
with respect to the coordinate $r$ with $q=2$, $Q=4$, $M=5$ and the largest radius putted at $r_{s}=11.38808027313$.}
\end{figure}

We have numerically found that the radius of the star is a continuous physical parameter
for the supporting star in a reflecting box,
whereas the radii of the supporting stars without reflecting boxes are discrete.
We mention that the box reflecting boundary and the star reflecting surface can play
the role of the infinity effective potential well
to confine the scalar field.
It seems that the potential well can make the scalar field easier
to condense and thus the radii of hairy stars becomes continuous after imposing a box boundary.
In fact, scalar condensation due to the confinement was also observed in black holes.
For example, it has been proved in \cite{Hod-11,Hod-12} that
Reissner-Nordstr$\ddot{o}$m black holes are stable even under charged scalar perturbations
in accordance with the no-scalar-hair theorem of black holes and in contrast,
charged black holes in a box can dynamically evolve into quasi-local hairy black holes
with charged scalar perturbations \cite{Carlos,Dolan,Supakchai,Nicolas}.

For given set of parameters, we can search for the largest radius $R_{s}$
below the upper bound of (17). In Fig. 3, we show effects of the star charge
and mass on the largest radius with dimensionless quantities according to the symmetry (18).
In the left panel, we plot $\mu R_{s}$ as a function of $qQ$ with $qM=10$.
We can see from the left panel that larger $qQ$ corresponds to a larger $\mu R_{s}$
similar to cases in the flat spacetime limit \cite{Hod-8}.
With $qQ=8$ in the right panel, we see that the
larger star mass $qM$ leads to a larger radius $\mu R_{s}$.
We can also see that the radius is almost linear with the charge and mass of the star.

\begin{figure}[h]
\includegraphics[width=165pt]{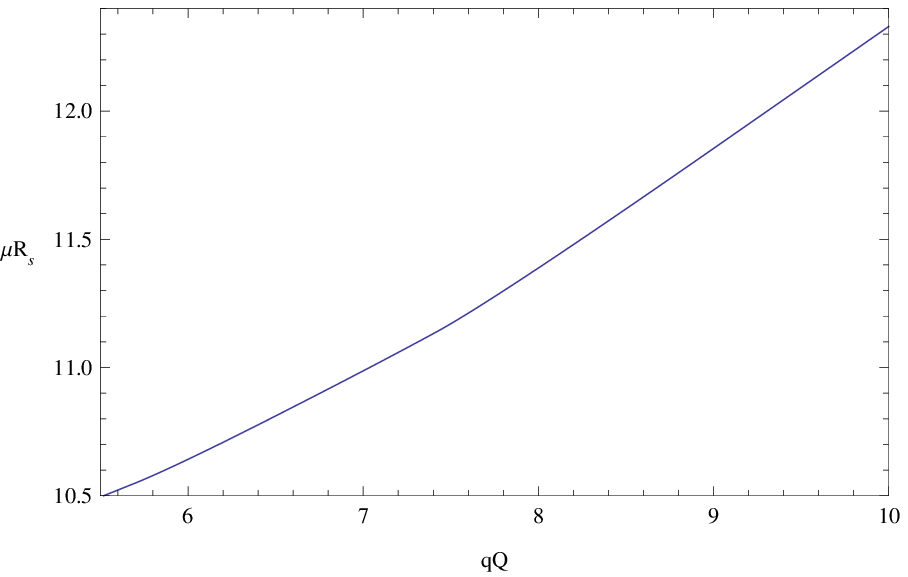}\
\includegraphics[width=165pt]{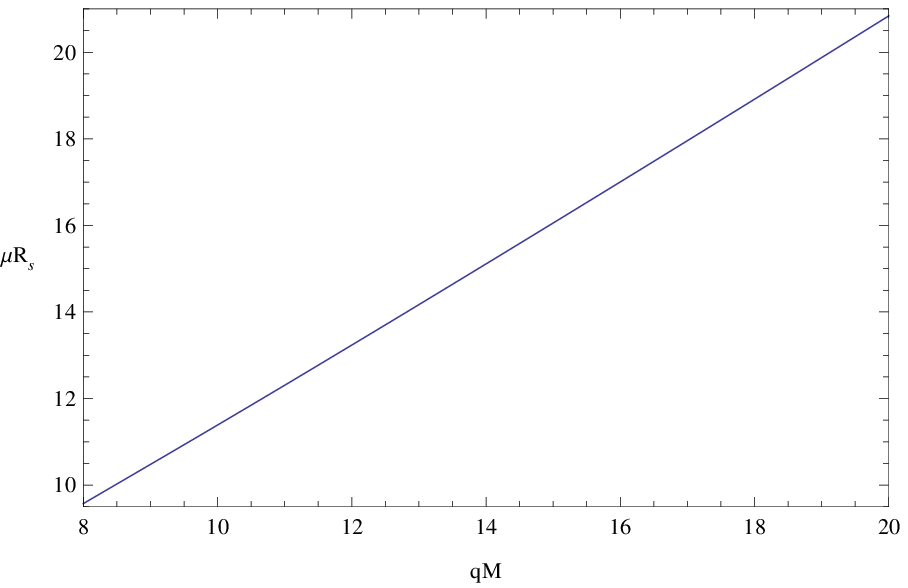}\
\caption{\label{EEntropySoliton} (Color online) We show the largest hairy star radius as a function of
the star charge and mass. The left panel shows behaviors
of $\mu R_{s}$ with respect to $qQ$ with $qM=10$
and the right panel represents effects of the star mass $qM$ on $\mu R_{s}$ with $qQ=8$.}
\end{figure}

\subsection{Analytical studies of scalar field configurations in the limit of flat spacetime}

In this part, we give an analytical treatment of the model
in the limit of $M,Q\ll r_{s}$ or the spacetime outside the star is flat.
In this case, the equation (5) of the scalar field can be putted as
\begin{eqnarray}\label{BHg}
r^2\tilde{\psi}''+r\tilde{\psi}'+(-\frac{1}{4}+q^2Q^2-\mu^2r^2)\tilde{\psi}=0,
\end{eqnarray}

Here, we still use the symmetry (18) to set $\mu=1$.
The general real solution of this radial differential equation can be expressed in terms of the modified Bessel functions \cite{Hod-8,Abramowitz}
\begin{eqnarray}\label{BHg}
\tilde{\psi}(r)=A\cdot K_{\nu}(r)+B\cdot I_{\nu}(r),
\end{eqnarray}
where $\nu^{2}=-(qQ)^2<0$ and A,B are constants.

According to boundary conditions (4), we have
\begin{eqnarray}\label{BHg}
A\cdot K_{\nu}(r_{s})+B\cdot I_{\nu}(r_{s})=0,
\end{eqnarray}
\begin{eqnarray}\label{BHg}
A\cdot K_{\nu}(r_{b})+B\cdot I_{\nu}(r_{b})=0,
\end{eqnarray}

In order to have a nontrivial scalar field $\tilde{\psi} \not \equiv0$, we have to impose that
\begin{eqnarray}\label{BHg}
H_{\nu}(r_{s},r_{b})=
\left|
  \begin{array}{cc}
K_{\nu}(r_{s}) & I_{\nu}(r_{s})\\
K_{\nu}(r_{b}) & I_{\nu}(r_{b})\\
  \end{array}
\right|
=0.
\end{eqnarray}

We can numerically search for the values $r_{s}$ and $r_{b}$ satisfying the equation (23). We plot
$H_{\nu}(r_{s},r)$ as a function of $r$ with $\nu=20i$ and different fixed values of $r_{s}$ in Fig. 4.
In the left panel with $r_{s}=15.34186$, we show that the box boundary can be putted
at $r_{b}=26.64315$. As we increase the value of $r_{s}$, $r_{b}$ becomes larger and
in the middle panel, when $r_{s}=15.34286$, $r_{b}$ is very far away from the star.
In this case, our quasi-local solution nearly goes back to cases without box boundary
and it is reasonable that the values $r_{s}=15.34286$ here is in good agreement
with those of hairy stars without box boundary in \cite{Hod-8}.
And in the right panel with $r_{s}=15.34386>15.34286$, we see that there is no place
to impose the reflecting box boundary.

According to (17), there are bounds for radius of hairy stars in the flat space limit
as $0< r_{s}\leqslant qQ=20$. We further show that the
radius of the hairy star is continuous in the range $(0,R_{s}]$ with $R_{\nu}\thickapprox15.34286$.
Here, we have included cases of $r_{b}=\infty$ with discrete hairy star radius around
$r_{s}\thickapprox15.34286,~~12.35398,~~10.21688, \cdots$. And for other $r_{s}$ between
the front discrete radius, we can find a hairy star with a corresponding box boundary at a fixed distance $r=r_{b}$.
In summary, we conclude that the radius of the hairy star in a box is continuous in the range of $(0,15.34286]$,
which is very different from cases without box boundaries where the radius of the hairy star is discrete in \cite{Hod-8,Hod-9,Hod-10}.

\begin{figure}[h]
\includegraphics[width=155pt]{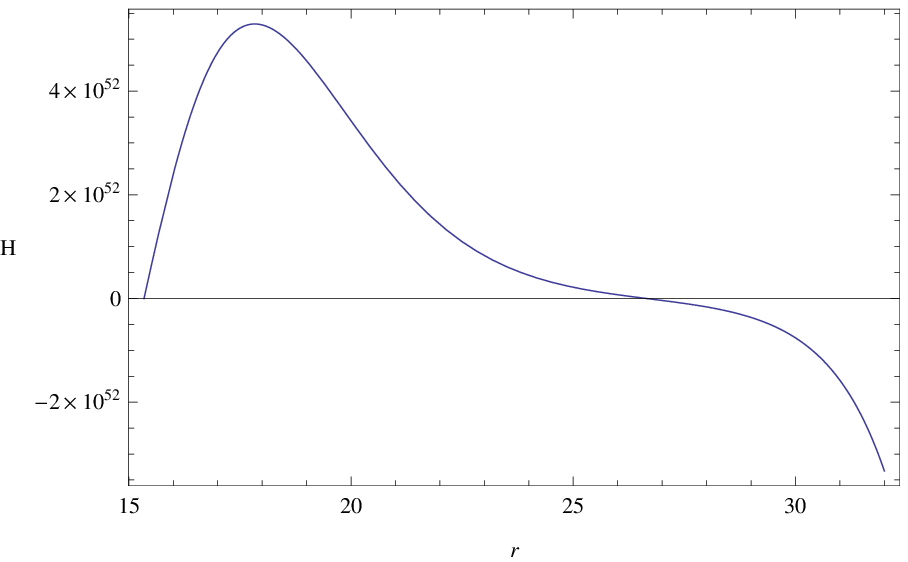}\
\includegraphics[width=155pt]{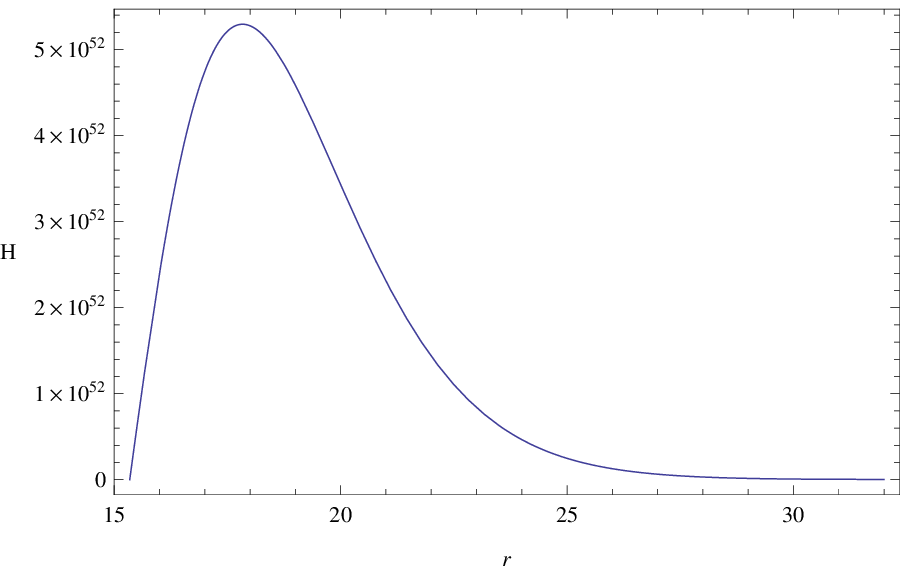}\
\includegraphics[width=155pt]{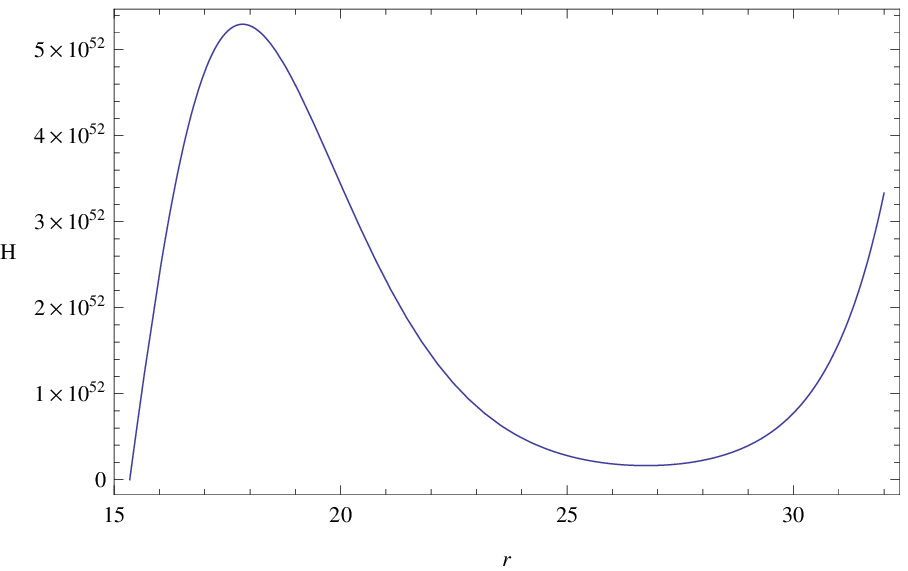}\
\caption{\label{EEntropySoliton} (Color online) We plot the function of $H_{\nu}(r_{s},r)$
with respect to the coordinate $r$ with $\nu=20i$. The left panel shows the case
of $r_{s}=15.34186$, the middle panel corresponds to the case of $r_{s}=15.34286$
and the right panel represents the case of $r_{s}=15.34386$.}
\end{figure}

\section{Conclusions}

We studied condensation of static scalar fields around charged compact reflecting stars
in the curved spacetime through both analytical and numerical methods.
We have enclosed the star in a box and our solutions can return to cases without box boundaries
when putting the box far from the star surface.
We focused on a scalar field of mass $\mu$ and charge coupling
constant q in the background of a compact reflecting star of radius $r_{s}$, mass M and electric charge Q.
We only researched on the case of $M\geqslant Q$ in this paper
and provided bounds for the radius of the scalar hairy compact star
as $\mu M+\sqrt{\mu^{2}(M^2-Q^2)} < \mu r_{s}\leqslant \mu M+\sqrt{\mu^{2}(M^2-Q^2)+q^2Q^2}$.
For certain set of parameters, we obtained numerical hairy reflecting star solutions satisfying boundary conditions.
In addition, we also analytically studied cases in the limit of $Q,M\ll r_{s}$.
With detailed calculations, we found that the radius of the hairy star is continuous
in a range with a corresponding box boundary at a proper distance and in contrast,
radius of hairy star without box boundaries is discrete in the range.
We also examined effects of the charge and mass of the star on the largest
hairy star radius $\mu R_{s}$ and found that larger star charge and star mass correspond to a larger $\mu R_{s}$.

\begin{acknowledgments}

We would like to thank the anonymous referee for the constructive suggestions to improve the manuscript.
This work was supported by the National Natural Science Foundation of China under Grant No. 11305097;
Shandong Provincial Natural Science Foundation of China.

\end{acknowledgments}

\end{document}